\documentclass[11pt]{article} 
\usepackage{amsfonts,amssymb,slashed,makeidx,latexsym,setspace}
\usepackage{graphicx,graphics,amssymb,epsf,rotate} 
\usepackage{times,cite,color}
\usepackage{hyperref}
\begin{document}
\begin{flushright}

\end{flushright}

\vskip 30pt

\begin{center}
{\Large \bf Gluino, wino and Higgsino-like particles without
  supersymmetry} \\
\vspace*{1cm} \renewcommand{\thefootnote}{\fnsymbol{footnote}} { {\sf
    Tirtha Sankar Ray, Hiroshi de Sandes, Carlos A. Savoy ${}$}} \\
\vspace{10pt} {\small ${}$ {\em \emph{Institut de Physique
      Th\'eorique, CEA-Saclay,\\ 91191 Gif-sur-Yvette,
      France}\footnote{Laboratoire de la Direction des Sciences de la
      Mati\`ere du CEA et Unit\'e de Recherche Associ\'e au CNRS
      (URA2306)}}}
\normalsize
\end{center}

\begin{abstract}
Metastable charged particles produced at the LHC can decay in the
quiescent period between beam crossing in the detector leading to
spectacular signals. In this paper we consider augmenting the Standard
Model with gluino, Higgsino and wino-like particles without invoking
supersymmetry. Proton stability is ensured by imposing a discrete
$Z_6$ parity that fixes their possible interaction and makes them
metastable in a large portion of the parameter space. We investigate
the possibility of employing these fields to improve gauge coupling
unification, explain dark matter, generate neutrino mass and cancel
flavor anomalies.  We find that the masses of these fields, controlled
by the flavor anomaly relations, make them visible at the LHC.
\end{abstract}

\section{Introduction}

If they exist, the LHC has the power to discover possible new fermions
beyond the TeV milestone, in particular, strongly interacting
ones. Many publications have discussed a variety of such states, their
signatures at colliders and, often, the models for physics beyond the
Standard Model (SM) that suggest their being.  A large amount of these
studies focus on heavy quarks and leptons in fundamental multiplets of
the SM. On the other hand, the supersymmetric extensions of the SM
generically require gauginos, in the adjoint representations of the SM
gauge group, as well as Higgsinos, electroweak doublets without lepton
number.  The corresponding interactions are supersymmetric transforms
of the gauge ones, and of the Higgs couplings, respectively. With
their signatures so fixed, these are currently the most hunted
particles at the LHC except, of course, for the Higgs.  It goes
without saying that they are instrumental for the beautiful
unification of the gauge couplings allowed in these models.

This paper discusses fermions with the same SM quantum numbers as
gauginos and Higgsinos in the absence of supersymmetry. Without
squarks and sleptons, they cannot mimic the supersymmetric couplings
and, besides the universal gauge interactions, have only effective
four fermion couplings to the standard fermions. Since they transform
in real representations of the SM symmetries, their masses are
expected to be near the cut-off scale, unless they are protected by
some mechanism. It seems plausible to take advantage of the flavor
symmetry that would allow for the quarks and leptons to have masses
way below the electroweak scale to suppress the mass of these exotic
states. In particular, if that symmetry is gauged, they could usefully
contribute to the cancellation of anomalies generated by the SM
fermions. As argued below, this scenario makes sense and allows for
new fermions within the LHC reach, in spite of the relatively high
limit on the flavor symmetry breaking scale imposed by data on FCNC
and CP violating effects.

Indeed, the strong hierarchy in the measured fermion masses and mixing
remains a mystery till date.  An elegant resolution of this problem is
achieved in the framework of Froggatt-Nielsen mechanism
\cite{Froggatt:1978nt}.  In the simplest form this entails the
prediction of an additional $U(1)_X$ flavor symmetry group, under
which the SM fermions are charged. This group, which may be gauged, is
spontaneously broken at some high scale by the vev of the flavon
scalar ($\phi$), with a conventionally assigned flavor charge $-1$.
At the weak scale, the coefficient of an SM operator ($\hat{O}$) is
suppressed by a factor $\epsilon^{|\Delta X|}$, where $\Delta X$ is
the $U(1)_X$ charge mismatch in the operator ($\hat{O}$) and
$\epsilon={\langle \phi \rangle}/{\Lambda}$, where $\langle \phi
\rangle$ is the flavon vev and $\Lambda$ is the cut-off. By applying
this rule to quark and lepton masses, including effective neutrino
mass operators, their measured patterns can be fairly reproduced by
suitably choosing the flavor charges.

It is reasonable to assume that at high energies, this effective
theory gives way to a renormalizable UV complete theory.  In this
context, the cancellation of chiral anomaly related to this $U(1)_X$
charge becomes a relevant issue.  Traditionally it has been assumed
that such anomaly cancellation takes place at the string scale through
the Green-Schwarz mechanism. Recently it has been suggested that such
anomaly cancellation can be achieved through additional exotic
fermions, some of which have masses in the TeV scale
\cite{Savoy:2010sj}, within the reach of the present generation
colliders. To facilitate anomaly cancellation, we analyze the
possibility of augmenting the SM with fermions in the adjoint
representation of the SM gauge group. We find that the addition of the
adjoint fields $\tilde{g}(8,1,0)$ and $\tilde{w}(1,3,0),$ and
$\tilde{h_d}(1,2,-\frac{1}{2})$, $\tilde{h_u}(1,2,+\frac{1}{2})$
represent the minimal content that can, in principle, cancel all the
anomalies in the theory if they are endowed with flavor
charges\footnote{Even if it could look inappropriate, for simplicity,
  we name and denote these new states just like the corresponding ones
  in the MSSM. }. We find that the cancellation of anomaly with this
minimal content implies large $U(1)_X$ charges for these fields. And
thus, in many instances, some of these fields have masses in the TeV
scale.

Interestingly, this also represents the fermionic sector of the
minimal SUSY extension of the SM with the exception of the hypercharge
gauge fermion, the bino. A similar SM singlet field charged only under
$U(1)_X$, would play no role in the cancellation of mixed anomalies
and may be added to the model with obvious alterations.  To make the
identification more alike we further propose to expand the scalar
sector to include a second Higgs doublet.  We assume a $Z_2^H$
symmetry that allows one of the Higgs to give mass to the up-type and
the other to the down-type. In principle the Higgs fields can be
charged under the $U(1)_X$ gauge group.  When the charges assigned to
the various fermions in the model are not integers, the flavon breaks
$U(1)_X$ into a discrete symmetry that plays an important role for the
consistency of the model. We assume it to be the so-called proton
hexality~\cite{Dreiner:2005rd} that forbids proton decay effective
operators. It also forbids mixing between the new and SM fermions,
even in presence of electroweak symmetry breaking, making the lightest
new fermion metastable, or even stable, hence a dark mater candidate.
In particular it forbids mixing between the Higgsino and the Lepton
analogous to the R parity introduced in the MSSM.
However it is consistent with lepton-wino mixing through the Higgs,
providing a model for type III leptogenesis.

Of course, a crucial difference is the absence of scalars other than
the Higgs fields, and of the corresponding dimension four
couplings. Therefore the lightest heavy fermion decays into three SM
states and becomes long-lived as the cut-off scale is high. In some
instances, the phenomenology of the models somewhat mimic the split
supersymmetry scenarios~\cite{ArkaniHamed:2004fb}, where a striking
feature is the possibility of observation of metastable
gluinos~\cite{ArkaniHamed:2004yi}.  In the present paper, we study the
possibility of having metastable fermions in the weak scale that can
be stopped in the detector, where they decay in a non-standard way at
a later instant.  We find that the decay pattern of the fields
considered in this paper are distinct from the gluinos from the split
SUSY framework and can be easily distinguished.

We also note that both the CMS and the ATLAS experiments at LHC
\cite{Khachatryan:2010uf,Aad:2011hz}
have searched for these metastable states and put initial mass limits
of the fields. These limits do not severely constrain this class of
models at present. However they can potentially be probed in the near
future with modified search strategy.

Since this is an alternative to the supersymmetric theory discussed in
\cite{Savoy:2010sj}, we refer to that paper for several details and
issues that are not reproduced here. However many relations are very
different because the gauginos in \cite{Savoy:2010sj} are the real
thing, and the heavy states are leptons and quarks and their scalar
companions. A recent paper \cite{Eboli:2011hr} builds
non-supersymmetric versions with metastable exotic quarks and leptons.

\section{The Minimal and Next to Minimal Models}
Although our approach should be rather generic in the framework of
gauged flavor symmetry, it will be presented here in simple cases by
way of illustration. In our minimal model, we consider the SM together
with the Froggatt- Nielsen model described in the introduction,
augmented with adjoint fermions and two weak doublets to compensate
for the anomalies related to a gauged abelian flavor symmetry. We also
enlarge the scalar sector to include an extra Higgs doublet. This
eases anomaly compensation.  These exotic fields are quite different
from the supersymmetric gauginos and Higgsinos.  For example they have
non trivial charges under the flavor symmetry, which among others
prevent the SU(2) adjoint triplet from mixing with the SU(2) doublets,
and, most importantly, they have very different couplings to quarks
and leptons. As discussed below, no simple example with unification of
gauge couplings was found for this minimal model.  In our next to
minimal model, an $I=0$ exotic charged lepton ($E$) is added.  This
gives some additional renormalization of the hypercharge coupling
improving among other things the possibility of driving gauge coupling
unification around the usual GUT scale.

These ersatz particles interact with the lighter SM fields essentially through
effective four fermion operators. It is well known that if higher
dimensional operators are added to the SM, it predictably leads to the
pitfall of rapid proton decay. This means that the cut-off of the
effective higher dimensional operators are essentially pushed to the
GUT scale. This situation can be evaded by considering additional
symmetry in the theory that prevents operators responsible for proton
decay from showing up.  In the present context this can be achieved by
simply considering a $Z_6$ symmetry, proton hexality~
\cite{Dreiner:2005rd}, under which the SM fields are suitably
charged. One can embed this discrete group into the continuous
symmetry group $U(1)_X$ so that when the flavon gets a
vev the flavor symmetry is spontaneously broken to the proton hexality
subgroup.

For each fermion f, one can separate the integer and fractional part,
$X_f =~x_f + Z'_f$, where $x$ is an integer. The fractional part of
the charges remains operative below the electroweak symmetry breaking
scale and prevents potentially dangerous operators in the theory. The
corresponding charge assignment of the SM fields are shown in
Table~\ref{table1}. The Higgs fields are assumed to have integer flavor
charges\footnote{Actually the consideration of consistent anomaly cancellation also
leads to a zero fractional flavor charge for the Higgs, as was argued in \cite{Savoy:2010sj}
and \cite{Dreiner:2005rd}.} to protect the discrete symmetry below the EWSB scale. 
 The exotic fields introduced are also charged
under this discrete symmetry. The choice of these charges is almost
unique as it determines the possible four fermion interaction of the
exotic fields and therefore their decay patterns. We exhibit the
possible choices in Table~\ref{table2} consistent with Majorana masses
for gauginos.

The theory is defined by the most general effective Lagrangian
consistent with the SM symmetries, proton hexality and a cut-off
$\Lambda$. As already noted, the various operators are modulated by
the Froggatt-Nielsen factors, i.e., powers of $\epsilon$ equal to
their flavor charges.  The lowest dimension operators include the
Higgs couplings to fermions that yield their masses and mixing and the
four fermion interactions. The lowest dimension operators that lead to
FCNC put severe bounds on the cut-off scale, as we now turn to
discuss. They are also responsible for the masses of the heavy
fermions and their couplings to the light ones, as will be discussed
later.
\begin{center}
\begin{table}
\begin{minipage}[t]{0.47\textwidth}
\begin{tabular}{|c|c|c|c|c|} \hline
&$SU(3)_c$&$SU(2)_L$&$Y$&$z^{\prime} \times 18$\\ \hline
  $q$&3&2&$1/6$&1\\ \hline $u$&$\bar{3}$&1&$-2/3$&-1\\ \hline
  $d$&$\bar{3}$&1&$1/3$&-1\\ \hline $l$&1&2&$-1/2$&-9\\ \hline
  $e$&1&1&1&9\\ \hline
\end{tabular}
\caption{\small \sf  
Charges of SM  fields, $z^{\prime} = z + \frac{1}{3} Y$}
\label{table1}
\end{minipage}
\begin{minipage}[t]{0.47\textwidth}
\centering
\begin{tabular}{|c|c|c|c|c|} \hline
&$SU(3)_c$&$SU(2)_L$&$Y$&$z^{\prime} \times 18$\\  \hline
$\tilde{g}$&8&1&$0$&$  9$\\  \hline
$\tilde{w}$&$1$&2&0&$  9, 0$\\  \hline
$\tilde{h_d}$&1&2&-1&$\pm 3$\\   \hline
$\tilde{h_u}$&1&2&$1$&$\mp 3$\\   \hline \hline 
$E$&1&1&-1&-3 \\ \hline
\end{tabular} 
\caption{\small \sf  Charges of exotic fields.}
\label{table2}
\end{minipage}
\end{table}
\end{center}
By convenience, we shift the flavor charge by a fraction of the
hypercharge to cancel the charge of one of the Higgses ($H_u$) and
assign a charge $x_H$ to $H_d$. Let us define the combinations of
flavor charges associated to the Yukawa couplings of light quarks to
the two Higgs:
\begin{eqnarray}
\chi_u^{ij}= x_q^i + x_{\bar{u}}^j \, , \qquad \chi_d^{ij} = x_q^i +
x_{\bar{d}}^j + x_H \, , \qquad \chi_e^{i} = x_l^i + x_{\bar{e}}^i +
x_H \label{chis}.
 \end{eqnarray}
The mass matrices of the fermions are given by, $m^{ij}_f \sim
\epsilon^{|\chi_f^{ij}|}v_f$, where $v_f$ is the vev of the
corresponding Higgs field. \footnote{An obvious  limitation of the FN approach 
with an abelian flavor symmetry is that all quantities are defined up to 
$O(1)$ factors. Therefore we use the symbol "$\sim$" to express these uncertainties.} From the known quark and lepton masses and
mixing, one can find the values of these Yukawa charges as a function
of $tan\beta = v_u /  v_d \sim 
\epsilon^{-x_t}.$\footnote{Notice that the  $x_t$ parametrizes 
the effects of $\tan \beta$ on the determination of the $\chi$ matrices
from the fermion masses. Hence only the integer part of $x_t$ is relevant,
so that $x_t = 0,\ 1$ or 2.}
 For $\epsilon \sim .2$ a
reasonable mass matrix is obtained if we impose the following
constraints: $ \chi_u^{ii} =(8,4,0)$; $\chi_d^{ii} + x_t =(7,5,3) $
and $ \chi_{e}^{ii} \pm x_t.  =(\pm 7, \pm 5 , \pm 3 )$. Many other
examples can be found in the literature, see, {\it e.g.},
\cite{Ibanez:1994ig}.

The effective neutrino mass matrix can be generated after integrating
out the physics above the cut-off, which would result into a dimension
five operator $( L H_u)^2$ in the effective Lagrangian. If the wino
has half-integer flavor charge as the leptons, it can mediate Type III
seesaw~\cite{Foot:1988aq} through its coupling $\bar{\tilde{w}} H_uL$
to the lepton doublets.  Integrating out the wino with mass
$\epsilon^{ |\chi_{\tilde{w}}|}\Lambda$, where $\chi_{\tilde{w}} = 2
x_{\tilde{w}}$, one gets a second contribution to neutrino masses.  In
order to ensure a neutrino mass matrix with a modest hierarchy, one
assumes that the $x_l^i = x_l$ is the same for all three leptons.  The
order of magnitude of final neutrino masses are,
\begin{equation}
m_{\nu} \sim \left( \epsilon^{ |2x_l - 1|} +\epsilon^{|2x_l - 1+
  \chi_{\tilde{w}}| -|\chi_{\tilde{w}}|} \right) \frac{v_u^2}{\Lambda},
\label{ssint}
\end{equation}
which numerically requires a large cut-off or large charges and puts
obvious restrictions on the latter. Note that the wino contribution is
similar or larger than the effective dimension five operator $(L
H_u)^2$. Clearly, fake binos can be added to implement Type I
seesaw. They contribute only to the flavor boson anomalous
self-coupling.

A large lower limit on the scale $\Lambda$ arises from the comparison
of effective FCNC four fermion operators and the data on rare
processes \cite{Isidori:2010kg}, but most results depend on the
suppression of the coefficients by the flavor model. However, for
broken gauged flavor symmetry there is a general limit arising from
the exchange of the massive flavor gauge boson
\cite{ArkaniHamed:1999yy}. The FCNC and CPV effects appear in the
flavor current when it is transformed to the physical basis because
the fermion charges are different, which produces a mixing
pattern comparable to the CKM matrix. This is generic and quite model
independent, entailing a quite general limit on the flavor breaking
scale, $\Lambda > 5\times 10^4\,$TeV. Obviously our models have to
respect this bound, but then all their operators are consistent with
the FCNC and CPV data without further restrictions.

Let us now turn to the discussion of the anomalies and their
cancellation through the heavy states. They are extensively discussed
in the literature \cite{Ibanez:1994ig} within the SM or the MSSM fermion
content. Three are linear in the flavor charges, the fourth is
quadratic. We display their expressions in a convenient reshuffled
form as follows:
\begin{eqnarray}
3 \chi_{\tilde{g}} =Tr(\chi_u + \chi_d) \qquad \chi_{\tilde{w}}+ \chi_{\tilde{h}} - 4 \chi_{\tilde{g}}
+\chi_E = -Tr(\chi_l - \chi_d) \ , \qquad 3 \sum_i {x^i}_q + 3 x_l + 2
\chi_{\tilde{w}} + \chi_{\tilde{h}} =1 \, ,&& \nonumber \\ \chi_{\tilde{h}} (x_{{\tilde{h}}_u} - x_{{\tilde{h}}_d}) -
\chi_E (x_E -x_{\bar{E}}) - {\cal F} =\sum_i [2\chi^i_u (x^i_q -
  x^i_u) - \chi^i_d (x^i_q - x^i_d) - \chi^i_l (x^i_l - x^i_e)] \, ,
\ \ \ \ \ \ \ \ && \nonumber \\ {\cal F} =\frac{1}{9}[Tr \{2 \chi_u -
  \chi_d + 9 \chi_l\}\pm 3 \chi_{\tilde{h}}. \qquad \chi_{\tilde{g}} = 2x_{\tilde{g}} \, , \quad
  \chi_{\tilde{w}} = 2x_{\tilde{w}}\, , \quad \chi_{\tilde{h}} = x_{\tilde{h_u}} + x_{\tilde{h_d}}\, , \quad \chi_E
  = x_E +x_{\bar{E}}\, .&&
\end{eqnarray}

The contribution of the exotic heavy lepton $E$ which is not present
in the minimal model is also shown. It is easy to realize from these
equations what is the minimal set of new fermions to compensate for
the anomalies, and why, besides improving gauge coupling unification,
it helps solving the anomaly equations. Actually, in the minimal model
many solutions to these equations imply a very light particle, already
excluded by experiments.  Some solutions -of course, there are many -
are shown in Tables \ref{minimaltb} and \ref{nminimaltb} to illustrate
various kinds of scenarios, which we now turn to discuss.

\section{Ersatz gluinos, winos and Higgsinos}
Since they transform in real representations of the SM gauge group,
the heavy fermions are expected to get masses $O(\Lambda)$, suppressed
by Froggatt-Nielsen factors. Namely, $m_f \sim \epsilon^{|\chi|_f|}
\Lambda$, with $\Lambda > 5\times 10^4\,$TeV, and the exponents
$\chi_f$ just defined for the exotic fermions. We look for solutions
of the anomaly equations that predict states in the TeV region. With
the small set of anomaly compensator fermions considered here, this is
often the case once the solutions with too light states are
discarded.

Besides these masses generated at the cut-off scale, Higgs couplings
can give rise to mass terms analogous to the SM ones. This happens in
particular, in the next-to-minimal model, for the system ($\tilde{h,}\,
E,\, H$).  Then one has the mass terms (in a sketchy notation):
\begin{equation}
a\epsilon^{ |\chi_E| }\Lambda \bar{E}E + b\epsilon^{
  |\chi_{\tilde{h}}| }\Lambda \tilde{h}_u \tilde{h}_d + c\epsilon^{
  |x_{h_u} + x_{\bar{E}}| } \langle H_u \rangle \tilde{h}_u \bar{E} + d\epsilon^{
  |x_{h_d} + x_{E}| } \langle H_d \rangle \tilde{h}_d E,
\end{equation}
with the explicit expressions for $m_E$, $m_{\tilde{h}}$, $m_+$ and
$m_-$, respectively, where $a,b,c,d$ are all $O(1)$. Because these
states have not been observed, one expects $m_{E, \tilde{h}} \gtrsim
v_u$. 

Note that for fermions with masses beyond 1~TeV, the contribution to
the oblique electroweak observables becomes negligible.  However the
contributions of the lighter states do need some attention. Indeed,
the couplings $(E{\tilde{h}}H)$ can be potentially dangerous if $E$
and $\tilde{h}$ are the lightest exotic states. In this case we
compute the contribution to the $T$ parameters using the mass
insertion approximation.  We find at the leading order, with $\xi= 1 -
m_E/ m_{\tilde{h}}$,
\begin{equation}
 \Delta T = \frac{1}{\sin^2\theta_W
   16\pi^2}\frac{m_-^2m_+^2}{M_W^2m_{\tilde{h}}^2} I(\xi),
\end{equation}
\footnotesize{\begin{equation} I(x)=\frac{x^{
        -4}}{3(x-2)^4}\left(4(x-2)x (x(156 +x(x(x
    +56)-158))-60)\right.  +\left.24(x-1)^2(20+x(x(24
    +x(x-12))-32))\ln(1-x)\right).
\end{equation}}\normalsize
The function varies between $I(x)\sim 1-8$ with $I(0)=6$.  With
conservative assumptions
$(|x_{{\tilde{h}}_d}+x_E|,|x_{{\tilde{h}}_u}+x_{E^{\prime}}|) \geq 1,$
we find that the theory passes the EWPT with ease and with practically
no constraints on the mass of the exotic fermions. As an illustration
we found that for $|\Delta T| < 0.1$ and
$(|x_{{\tilde{h}}_d}+x_E|,|x_{{\tilde{h}}_u}+x_{E^{\prime}}|)=1$ the
bound is given by $m_{\tilde{h}}>50~GeV.$ We have checked that all the
examples in Tables~\ref{minimaltb} and \ref{nminimaltb} pass EWPT.

The contributions due to the wino and Higgsino to the $T$ parameter in
the context of the MSSM were first computed in
\cite{Barbieri:1983wy}. They obtained an upper bound of $\Delta T <
0.09.$ Instead, with the $Z'$ chosen here, the two states cannot mix
through the Higgs coupling and do not contribute to $S$ and $T$.

As for their decays, at least the lightest ones must decay into four
fermions (or be stable and become a dark matter candidate). When these
particles get their masses around a few TeV, because the range of
their four fermion interactions is given by $\Lambda^{-1}$, they are
very long-lived at the scale of colliders, even without the
suppression due to Froggatt-Nielsen factors. The decay modes will
depend on the fermion structure of the interactions, the flavor
distribution of the rates being more model dependent.

With the charges as defined in Tables~\ref{table1} and \ref{table2},
it is easy to construct the possible four fermion operators that are
allowed by the discrete symmetry.  Considering the Majorana nature of
the fields, the only possibility for the gluino and wino is
$Z'=1/2$. For the Higgsino we also have two choices, $Z'(\tilde{h}_d)
=\pm3/18$. Below we summarize the possible decays:
 $$ \tilde{g}\rightarrow \bar{l} q \bar{u_R}, ~~ \bar{l} d_R
\bar{q},~~ \bar{e_R} \bar{u_R} d_R (+ \mathrm{h.c. })\qquad \qquad
\tilde{w}\rightarrow \bar{l} q \bar{u_R}, ~~ \bar{l} d_R \bar{q},~~
\bar{l} \bar{l} d_R, ~~ \bar{l} \bar{H_u} (+ \mathrm{h.c. }) $$
 $$ Z'(\tilde{h}_d)= -3/18: \ \tilde{h}_d \rightarrow
\bar{q}\bar{q}\bar{q}, ~~ \bar{q}\bar{u}\bar{d} \qquad
Z'(\tilde{h}_d)=3/18:\ \tilde{h}_d \rightarrow qdd \qquad \qquad E
\rightarrow \bar{q}\bar{q}\bar{u},~~\bar{d}\bar{u}\bar{u}\, .
$$ 
The family patterns of the decay products are very model
dependent. The lifetimes determined by these decay modes of the
lightest exotic particles, in some specific examples are given in the
tables.

The addition of adjoint and weak doublet fields has the potential to
drive gauge coupling unification. This can be anticipated from the
close resemblance of these models with the split SUSY scenario. Even
if exact unification is not achieved, in the majority of these models
a significant improvement in gauge coupling unification should be
possible. In order to make a quantitative study we define the
following parameters,
\begin{equation}
 \alpha_i(M_{\mathrm{GUT}}^{-1} - \alpha_i(M_Z)^{-1}= \Delta_i =
 \Delta_i^{\mathrm{SM}} + \Delta_i^{\mathrm{new}},
\end{equation} 
where $i$ corresponds to the three gauge groups of the SM and $
\Delta_i^{\mathrm{SM}}$ and $\Delta_i^{\mathrm{new}}$ are the
contributions from the SM (new) states , respectively, to the running
of the gauge couplings from $M_Z$ to $M_{\mathrm{GUT}}$.  Using the
experimentally measured low energy values of the gauge couplings, the
condition for gauge coupling unification in a generic model reduces to
the following expressions \cite{Nath:2006ut},
\begin{equation}
\rho = \frac{\Delta_2-\Delta_3}{\Delta_1-\Delta_2} = 0.719 \pm 0.005\,
, \qquad \qquad \Delta_1-\Delta_2 = 29.42 \pm 0.03. \label{unif}
\end{equation}
To analyze the improvement in gauge coupling unification
quantitatively we device the following strategy: for every model a
$M_{\mathrm{GUT}}$ value is defined from the $\Delta_1-\Delta_2$ given
in Eqs.~\ref{unif}. Then the following parameter is computed for the
new model,
\begin{equation}
 r = \frac{| \rho_{\mathrm{new}} -0.719|}{|\rho_{\mathrm{SM}}
   -0.719|}.
\end{equation}
In Tables~\ref{minimaltb} and \ref{nminimaltb} this parameter is
quoted for the specific models and it is also indicated when the
unification is consistent with the present experimental errors.

\begin{figure*}
\begin{center}
\includegraphics[width=.7\textwidth,keepaspectratio]
{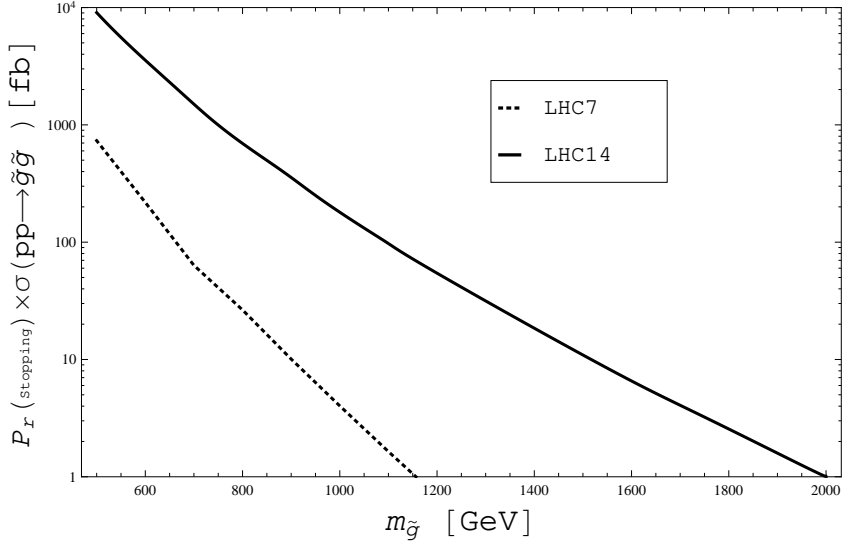}
\end{center}
\caption[]{\em \small The stopping cross-section for the gluinos at LHC.}
\label{fig1}
\end{figure*}

\section{Experimental consequences}
The exotic fermions introduced here, have four fermion interactions
that are determined by the ad hoc but motivated assignments of the
$Z'$ charges as given in Table~\ref{table2}. Deviation from these
specific choices can summarily kill all possible interaction leaving
these fermions stable at the cosmological scale and making them
possible dark matter candidates. For the Majorana fermions the choices
for the $Z'$ assignment are restricted to $\pm1/2$ or $0$. An
interesting scenario arises when $\chi_{\tilde{\omega}}$ is even. In
this case the $Z'$ charge for the exotic wino is zero and therefore it
cannot couple to the SM fermions. If this is the lightest exotic
fermion with mass around $\sim 2.7 ~TeV$ it can be a viable Minimal
Dark Matter (MDM) candidate\cite{Cirelli:2005uq} that satisfies the 7
year WMAP constraints \cite{Komatsu:2010fb}. However the nuclear
 cross-sections of these species get an unavoidable Sommerfeld
enhancement putting them in imminent danger of being ruled out by
indirect searches, specially by the anti-proton data
\cite{Cirelli:2008id}. At present the experimental observations are
subject to large astrophysical uncertainties leaving enough room for
the wino to survive as a DM candidate.

Another interesting scenario would be to consider the $Z'$ charge of
exotic Higgsino chosen to make it stable. It is known that a pure
Higgsino like stable particle can satisfy the WMAP data if it has a
mass around $\sim 1~TeV$ \cite{Giudice:2004tc}. Unfortunately the
coupling of the Higgsino to the $Z$ boson makes it unfavorable from
indirect searches.

The exotic fermions introduced can be within the experimental reach of
the LHC as is evident from Tables~\ref{minimaltb} and
\ref{nminimaltb}.  These exotic particles can be produced at the
collider and all of them are metastable except the exotic wino which
decays easily via the interaction $\tilde{w}H_uL$.  The
signals of the SU(2) adjoint triplet at the colliders closely resemble
the scenario of generic Type III seesaw models.  The triplet is
produced by gauge interactions $q\bar{q}\rightarrow
\tilde{w}^0\tilde{w}^0, \tilde{w}^+\tilde{w}^-$ and $u\bar{d}
\rightarrow \tilde{w}^+\tilde{w}^0 $. The latter is expected to have
the highest cross-section at the LHC which is $\sim 35 fb$ for
$m_{\tilde{w}} = 500 GeV$ and it falls to $\sim 1 fb$ when the mass
reaches $1 TeV.$ Once produced they mainly decay by $\tilde{w}^0
\rightarrow \nu h$ and $\tilde{w}^{\pm} \rightarrow l^{\pm} h.$ For
the range of parameters where Eq.~\ref{ssint} gives the correct
neutrino mass one expects a displaced vertex.  See
\cite{Franceschini:2008pz} for a detailed study of signals in Type III
seesaw models at LHC, that closely resemble the scenario where the
exotic wino is the only observable particle in the model.

Consider a scenario where the long living exotic gluino is the
lightest particle within the reach of the LHC. In this case the exotic
gluinos can be copiously produced at the hadron collider. Once
produced these particles will hadronize into R-hadrons and
R-mesons. They will lose considerable amount of energy as they travel
through the detector. A fraction of these hadronized exotic gluinos
will come to rest within the detector and then decay at a later
instance \cite{Arvanitaki:2005nq} giving rise to an interesting
signature.  Both the CMS \cite{Khachatryan:2010uf}
and ATLAS \cite{Aad:2011hz} have already published their
results on long living gluinos. The present bound on the particle
masses is around $ 400~GeV$, however we note that most of these
studies were carried out within the framework of the split SUSY
scenario.  In \cite{Khachatryan:2010uf} the probability for a produced
gluino to be stopped within the CMS detector was simulated.  In
Figure~\ref{fig1} we show the cross-section of stopped gluinos as a
function of gluino mass.  We have used CalcHEP 2.3.5
\cite{Pukhov:2004ca} to calculate the cross-section. Once stopped
these metastable particles will decay generally in the quiescent
period, i.e., out of sync with the proton-proton collision at the
collider. The decay will lead to a signal like: $1~prompt~lepton + 2~
jets$. This decay is different from the ones studied at the LHC
experiments within the split SUSY paradigm. We note that the prompt
lepton in the decay of the exotic gluino in this class of models
clearly distinguishes it from the split SUSY signals.

The production cross-section for the exotic Higgsino and the exotic
lepton is much smaller than that of the exotic gluino. Nevertheless
their decay after being captured in the detector can give rise to
spectacular signals that can easily be identified. However it is their
slow production rates that makes it difficult to observe at low
energy/luminosity.

\section{Conclusion}
In this paper, we have studied the properties of particles with the
quantum numbers of the sfermions of supersymmetric versions of the SM
in the total absence of supersymmetry.  We have shown that they can
play a role in gauged flavor symmetry to cancel the anomalies due to
the quarks and leptons. A crucial point is the discrete symmetry that
survives flavor symmetry breaking - proton hexality - to stabilize the
flavor models (and the proton!). In most of the viable examples of
this framework, some of the new states might appear within the reach
of the LHC as metastable particles with characteristic decay
patterns. Alternatively, the set up encompasses a version of the Type
III seesaw phenomenology, with a weak isospin triplet fermion whose
mass is light because of the proposed mechanism. Another consistent
possibility is that this triplet is stable and is a dark matter
candidate.  This scenario can be easily extended to non-abelian flavor
symmetry.

\vskip 2pt

\noindent {\bf{Acknowledgments:}}~ We thank Marco Cirelli for
discussions. This work has been partially supported by the Agence
Nationale de la Recherche under contract ANR 2010 BLANC 041301. The
work of TSR is supported by EU ITN , contract "UNILHC"
PITN-GA-2009-237920, the CEA-Eurotalents program.  HS is supported by a
CAPES Foundation (Ministry of Education of Brazil) Postdoc Fellowship.

\small{
\begin{center}
\begin{table}
\centering
\begin{tabular}{|c|c|c|c|c|c|} \hline
\multicolumn{6}{|c|}{Minimal Model} \\ \hline
\multicolumn{2}{|c|}{Model Parameters}&$Exotic ~fields$&$Lightest ~particle$&$r$&$Decay ~width$ \\ \cline{1-2}
$Sl. No.$&$SM ~fields$&&&&\\ \hline
&$x_q=(2,0,-2)$&&&& \\ 
&$x_{\bar{u}}=(6,4,2)$&$\chi_{\tilde{g}}=-11$&&&\\
1&$x_{\bar{d}}=(7,7,7)$&$\chi_{\tilde{w}}=-1$&$ m_{\tilde{g}}\sim 1.5 ~TeV$&0.33&$C\tau(\tilde{{g}_1})\sim 5.6 \times 10^{12} ~km$\\ 
&$x_t=0,~ x_H=-2$&$\chi_{\tilde{h}}=-7$&&&\\
&$x_l=4, ~z^{\prime}(\tilde{h}_d)=-\frac{3}{18}$&&&&\\
&$x_{\bar{e}}=(-9,3,-1)$&&&&\\ \hline
&$x_q=(7,5,3)$&&&& \\ 
&$x_{\bar{u}}=(1,-1,-3)$&$\chi_{\tilde{g}}=-3$&&&\\
2&$x_{\bar{d}}=(6,6,6)$&$\chi_{\tilde{w}}=-11$&$m_{\tilde{w}}\sim 0.5 ~TeV$&1.48&$\tilde{w}$ is unstable\\ 
&$x_t=1, ~x_H=5$&$\chi_{\tilde{h}}=9$&&&\\
&$x_l=-3, ~z^{\prime}(\tilde{h}_d)=-\frac{3}{18}$&&&&\\
&$x_{\bar{e}}=(4,2,-4)$&&&&\\ \hline
&$x_q=(2,0,-2)$&&&& \\ 
&$x_{\bar{u}}=(6,4,2)$&$\chi_{\tilde{g}}=-3$&&&$C\tau(\tilde{h}_d)\sim 8.2 \times 10^{6}~ km$\\
3&$x_{\bar{d}}=(-1,-1,-1)$&$\chi_{\tilde{w}}=1$&$m_{\tilde{h}}\sim 0.5 ~TeV$&0.33&$C\tau(\tilde{h}_u)\sim 4.4 \times 10^{8}~ km$\\ 
&$x_t=0, ~x_H=6$&$\chi_{\tilde{h}}=9$&&&\\
&$x_l=-4,~ z^{\prime}(\tilde{h}_d)=-\frac{3}{18}$&&&&\\
&$x_{\bar{e}}=(5,-7,-5)$&&&&\\ \hline
&$x_q=(5,3,1)$&&&& \\ 
&$x_{\bar{u}}=(3,1,-1)$&$\chi_{\tilde{g}}=-5$&&&\\
4&$x_{\bar{d}}=(-2,-2,-2)$&$\chi_{\tilde{w}}=-10$&$m_{\tilde{w}}\sim 1.5 ~TeV$&0.86& $\tilde{w}$ is a dark matter candidate\\ 
&$x_t=2, ~x_H=2$&$\chi_{\tilde{h}}=2$&&&\\
&$x_l=4,~ z^{\prime}(\tilde{h}_d)=-\frac{3}{18}$&&&&\\
&$x_{\bar{e}}=(-1,-3,-7)$&&&&\\ \hline
\end{tabular}
\caption{\small \sf Examples for the minimal model }
\label{minimaltb}
\end{table}
\end{center}
}
\begin{center}
\begin{table}
\small{
\centering
\begin{tabular}{|c|c|c|c|c|c|} \hline
\multicolumn{6}{|c|}{Next to Minimal Model} \\ \hline
\multicolumn{2}{|c|}{Model Parameter}&$Exotic ~fields$&$Lightest ~particle$&$r$&$Decay ~width$ \\ \cline{1-2}
$Sl.No.$&$SM ~fields$&&&&\\ \hline
&$x_q=(4,2,0)$&&&& \\ 
&$x_{\bar{u}}=(4,2,0)$&$\chi_{\tilde{g}}=-5$&&&$C\tau(\tilde{h}_d)\sim7.5\times 10^7 ~km$\\
1&$x_{\bar{d}}=(-1,-1,-1)$&$\chi_{\tilde{w}}=-7$&$m_{\tilde{h}}\sim 1.5 ~TeV$&0.09&$C\tau(\tilde{h}_u)\sim1.8\times 10^9 ~km$\\ 
&$x_t=0,~ x_H=4$&$\chi_{\tilde{h}}=9$&&&\\
&$x_l=-4, ~z^{\prime}(\tilde{h}_d)=-\frac{3}{18}$&$\chi_E=2$&&Unification&\\
&$x_{\bar{e}}=(-7,-5,3)$&&&&\\ \hline
&$x_q=(6,4,2)$&&&& \\ 
&$x_{\bar{u}}=(2,0,-2)$&$\chi_{\tilde{g}}=-5$&&&\\
2&$x_{\bar{d}}=(-3,-3,-3)$&$\chi_{\tilde{w}}=-9$&$m_{\tilde{w}}\sim 1.5 ~TeV$&0.09&$\tilde{w}$ in unstable\\ 
&$x_t=2,~ x_H=2$&$\chi_{\tilde{h}}=-5$&&&\\
&$x_l=-4, ~z^{\prime}(\tilde{h}_d)=-\frac{3}{18}$&$\chi_E=4$&&Unification&\\
&$x_{\bar{e}}=(-3,5,3)$&&&&\\ \hline
&$x_q=(4,2,0)$&&&& \\ 
&$x_{\bar{u}}=(4,2,0)$&$\chi_{\tilde{g}}=-7$&&&\\
3&$x_{\bar{d}}=(1,1,1)$&$\chi_{\tilde{w}}=-10$&$m_{\tilde{w}} \sim 1.6 ~TeV$&0.05&$\tilde{w}$ is a dark matter candidate\\ 
&$x_t=2,~ x_H=0$&$\chi_{\tilde{h}}=-9$&&Unification&\\
&$x_l=4, ~z^{\prime}(\tilde{h}_d)=-\frac{3}{18}$&$\chi_E=1$&&&\\
&$x_{\bar{e}}=(-9,-1,-3)$&&&&\\ \hline
&$x_q=(-1,-3,-5)$&&&& \\ 
&$x_{\bar{u}}=(9,7,5)$&$\chi_{\tilde{g}}=-1$&&& $\tilde{w}$ is unstable\\
4&$x_{\bar{d}}=(0,0,0)$&$\chi_{\tilde{w}}=7$&$m_{E} \sim m_{\tilde{w}} \sim 0.5~TeV$&0.07&$C\tau({E})\sim 959~km$\\ 
&$x_t=0,~ x_H=8$&$\chi_{\tilde{h}}=-4$&&Unification&$C\tau(\bar{E})\sim 1.1 \times 10^9~km$\\
&$x_l=6, ~z^{\prime}(\tilde{h}_d)=-\frac{3}{18}$&$\chi_E=-7$&&&\\
&$x_{\bar{e}}=(-7,-9,-11)$&&&&\\ \hline
&$x_q=(5,3,1)$&&&& \\ 
&$x_{\bar{u}}=(3,1,-1)$&$\chi_{\tilde{g}}=-9$&&& \\
5&$x_{\bar{d}}=(2,2,2)$&$\chi_{\tilde{w}}=-3$& $m_{\tilde{g}} \sim 1.5~TeV$&0.9&$C\tau(\tilde{g})\sim 7.1 \times 10^6~km$ \\ 
&$x_t=0,~ x_H=0$&$\chi_{\tilde{h}}=-8$&&&\\
&$x_l=-4, ~z^{\prime}(\tilde{h}_d)=-\frac{3}{18}$&$\chi_E=5$&&&\\
&$x_{\bar{e}}=(-3,-1,1)$&&&&\\ \hline
\end{tabular}
\caption{\small \sf  Examples for the next to minimal model 
  }
\label{nminimaltb}
}
\end{table}
\end{center}

\end{document}